\documentclass[runningheads]{llncs}

\usepackage{graphicx}
\usepackage{rotating}
\usepackage{graphicx}
\usepackage{psfrag}
\usepackage{amsmath}
\usepackage{amssymb}

\usepackage{url}
\urldef{\mailsa}\path|mitrovic@phy.bg.ac.yu, Bosiljka.Tadic@ijs.si|
\newcommand{\keywords}[1]{\par\addvspace\baselineskip
\noindent\keywordname\enspace\ignorespaces#1}

\begin{document}

\mainmatter  

\title{Search of Weighted Subgraphs on Complex Networks with Maximum Likelihood Methods}

\titlerunning{Weighted Subgraphs}

%
%
\author{Marija Mitrovi\'c$^{1}$ \and Bosiljka Tadi\'c$^{2}$}
\authorrunning{Weighted Subgraphs}

\institute{
$^{1}$Institute of Physics, Belgrade, Serbia\\
 $^{2}$Jo\v zef Stefan Institute, Ljubljana, Slovenia\\
\mailsa \\
\url{http://scl.phy.bg.ac.yu, http://www-f1.ijs.si/~tadic}
}

%
%

\toctitle{Weighted Subgraphs}
\maketitle

\begin{abstract}
Real-data networks often appear to have strong modularity, or network-of-networks structure, in which subgraphs of various size and consistency occur. Finding the respective subgraph structure is of great importance, in particular for understanding the dynamics on these networks. Here we study modular networks using generalized method of maximum likelihood. We first demonstrate how the method works on computer-generated networks with the subgraphs of controlled connection strengths and clustering. We then implement the algorithm which is based on weights of links and show it's efficiency in finding weighted subgraphs on fully connected graph and on real-data network of yeast.
\end{abstract}

Published: \textit{Lecture Notes in Computer Science}, vol.5102, pages 551$-$558, 2008
\section{Introduction}

Complex dynamical systems can be adequately represented by networks with a diversity of structural and dynamical characteristics, \cite{SD_book}, \cite{Nets_review} and \cite{TRT07}. Often such networks appear to have multiscale structure with subgraphs  of different sizes and topological consistency. 
Some well known examples include gene modules on genetic networks \cite{genes}, social community structures \cite{BCN-emails}, topological clusters \cite{Internet} or dynamical aggregation on the Internet, to mention only a few. It has been understood that in the evolving networks some functional units may have emerged as modules or communities, that can be  topologically recognized by better or tighter connections.
Finding such substructures is therefore of great importance primarily for understanding network's evolution and function. 

In recent years great attention has been devoted to the problem of community structure in social and other networks, where the community is topologically defined as a subgraph of nodes with better connection among the members compared with the connections between the subgraphs, \cite{BCN-emails} and \cite{NG}. 
Variety of algorithms have been developed and tested, a comparative analysis of many such algorithms can be found in \cite{BCN-emails}. 
Mostly such algorithms are based on the theorem of maximal-flow--minimal-cut \cite{max-min-theorem}, where naturally, maximum topological flow falls on the links between the communities. 
Recently a new approach was proposed based on the {\it maximum-likelihood} method, \cite{MLM}.
In maximization likelihood method an assumed {\it mixture model} is fit to a given data
set. Assuming that the network nodes can be split into $G$  groups, where group
memberships are unknown, then the expectation-maximization algorithm is used in order
to find maximum of the likelihood that suites the model. As a result a set of probabilities
that a node belongs to a certain group are obtained. The probabilities
corresponding to global maximum of the likelihood are expected to give the best split of the
network into given number of  groups.  

In complex dynamical networks, however, other types of substructures may occur, that are not necessarily related to ``better connectivity'' measure. Generally, the substructures may be differentiable with respect certain functional (dynamical) constraints, such as limited path length (or cost), weighted subgraphs, or subgraphs that are synchronizable at a given time scale. Search for such types of substructures may require new algorithms adapted to the respective dynamical constraints. 

In this work we adapt the maximum-likelihood methods to study  subgraphs with weighted links in real and computer-generated networks. We first introduce a new model to generate a network-of-networks with a controlled subgraph structure and implement the algorithm, \cite{MLM}, to test its limits and ability to find the {\it a priori} known substructures. We then generalize the algorithm to incorporate the weights on the links and apply it to find the weighted subgraphs on a almost fully connected random graph with known weighted subgraphs and on a real-data network of yeast gene-expression correlations.

\section{Network of Networks: Growth Algorithm and Structure}\label{section:nets} 

We introduce an algorithm for network growth with a controlled modularity. As a basis, we use the model with preferential attachment and preferential rewiring, \cite{BT}, which captures  the statistical features of the  World Wide Web. Two  parameters ${\tilde{\alpha}}$ and $\alpha$ control the emergent structure of the Webgraph when the average number of links per node $M$ is fixed. For instance, for $M=1$: when ${\tilde{\alpha}} < 1$ the emergent structure is a scale-free clustered and correlated network, in particular the case ${\tilde{\alpha}}=\alpha =1/4$ corresponds to the properties measured in the WWW \cite{BT}; when ${\tilde{\alpha}} = 1$  a scale-free tree structure emerges with the exponents depending on the parameter $\alpha$. Here we generalize the model in a nontrivial manner to permit development of distinct subnetworks or modules. The number of different groups of nodes is controlled by additional parameter $P_{o}$. Each subgroup evolves according to the  rules  of Webgraph.
  
\begin{figure}
\centering
\begin{tabular}{cc} 
\includegraphics[height=0.25\textheight,width=0.3\textheight]{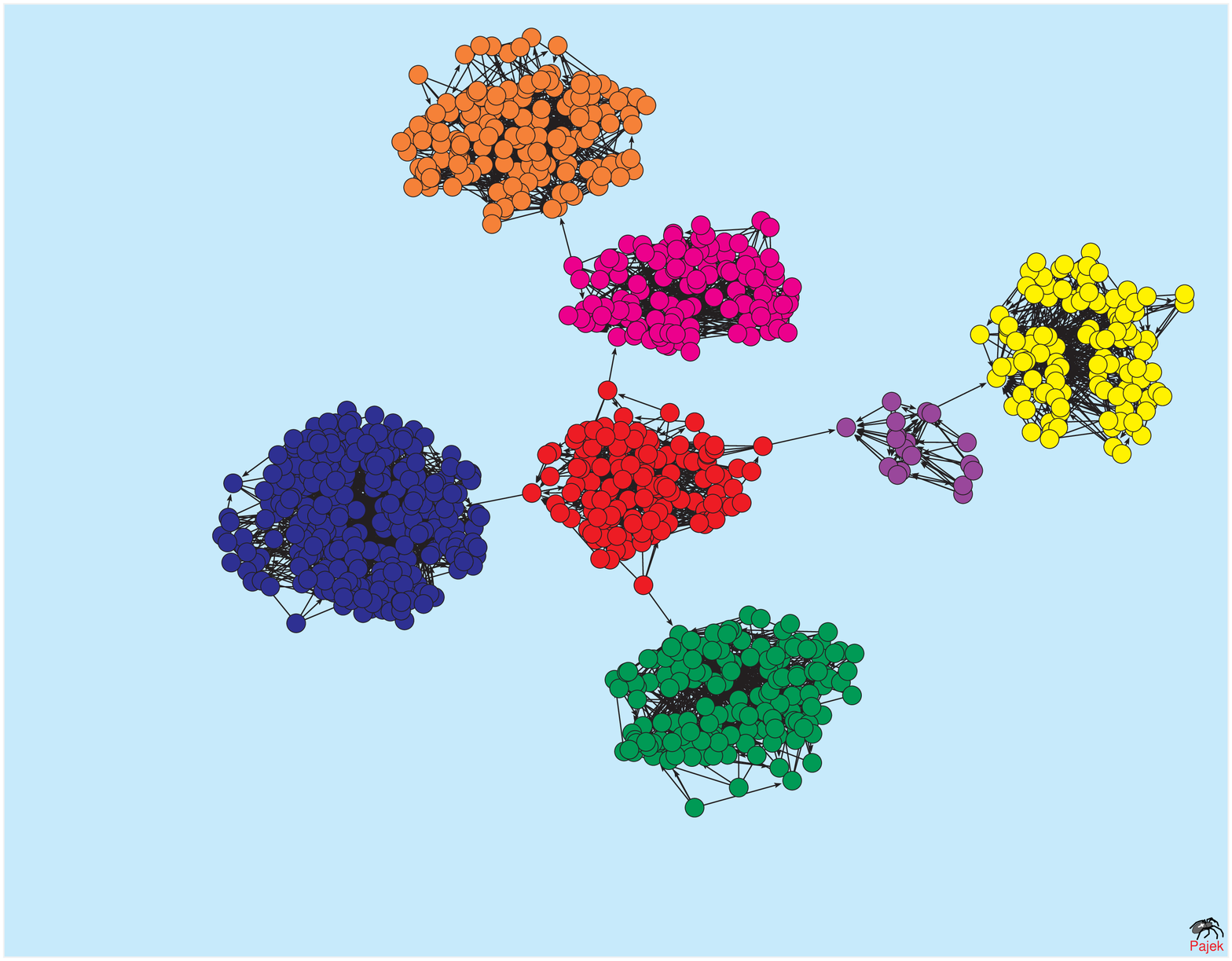}&
\includegraphics[height=0.25\textheight,width=0.3\textheight]{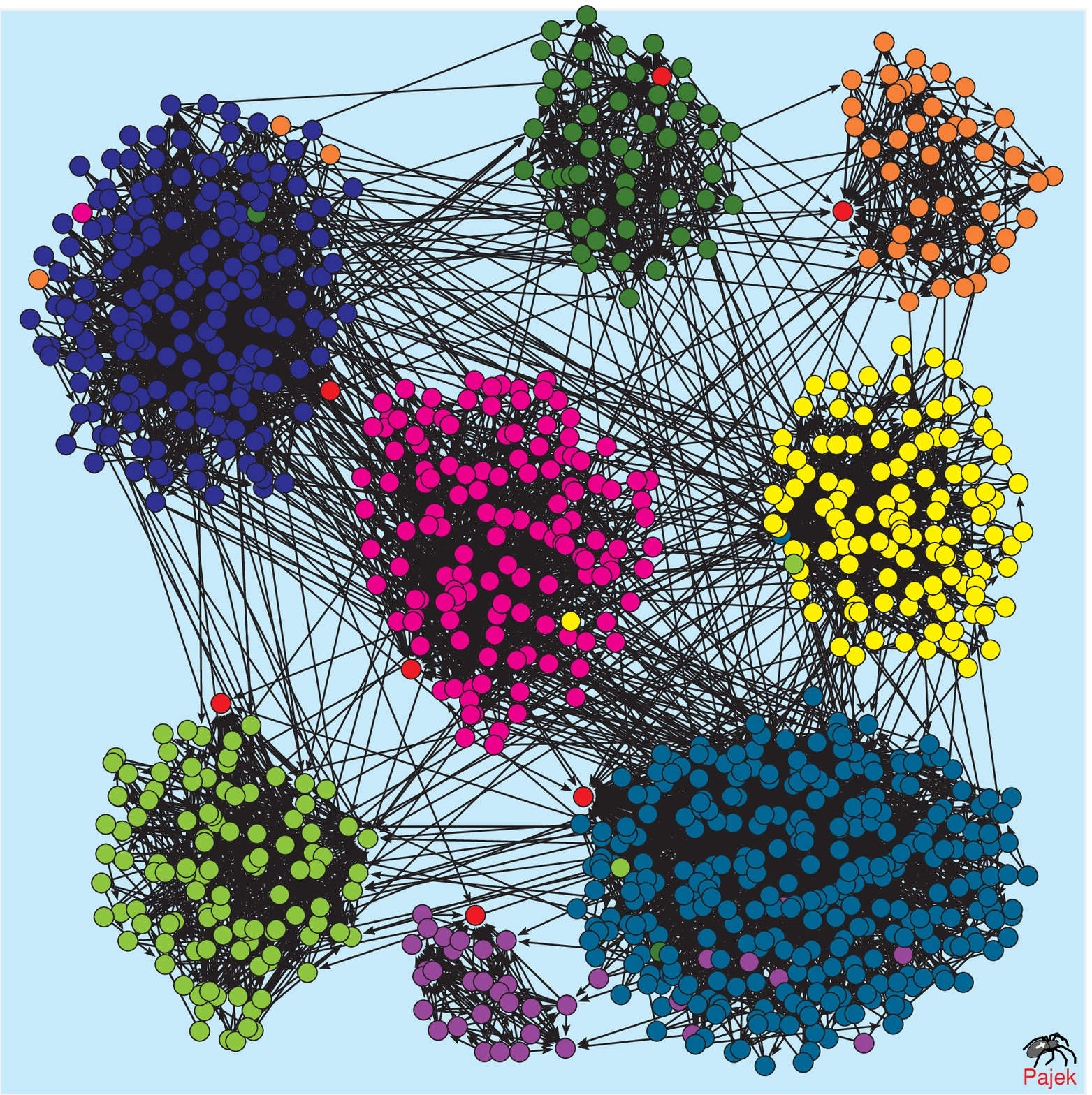}\\
\end{tabular}
\caption{Network of networks generated by the algorithm described above for $N=1000$ nodes and different combinations of control parameters $\alpha$,${\tilde{\alpha}}$ , $P_{o}$ and $M$. Different colors represent topological subgraphs as found by the maximum-likelihood method.}
\label{fig-graphs-clusters}
\end{figure}

\begin{figure}
\centering
\hskip -0.4 cm
\begin{tabular}{cc} 
\resizebox{14pc}{!}{\includegraphics{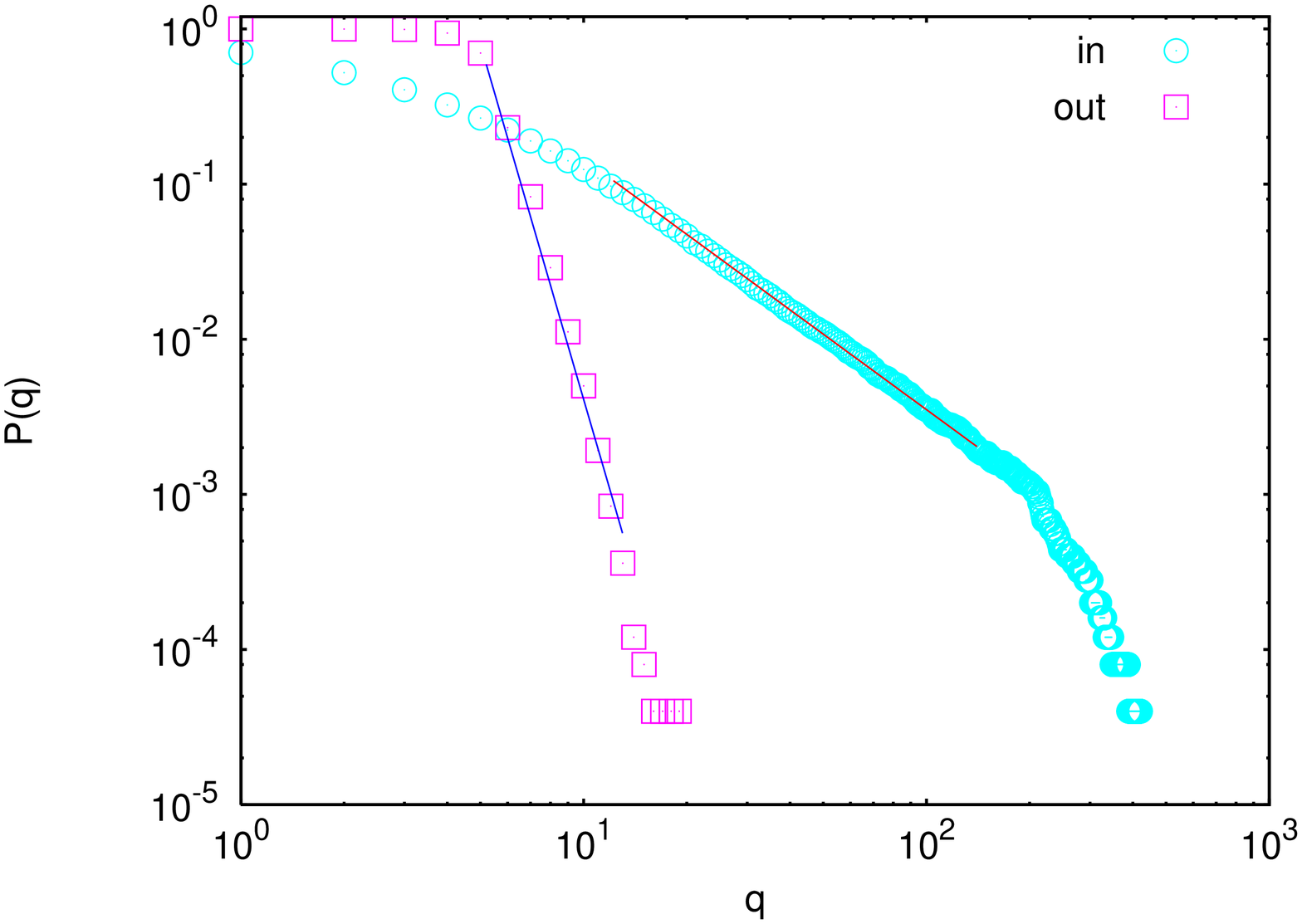}}& 
\resizebox{14pc}{!}{\includegraphics{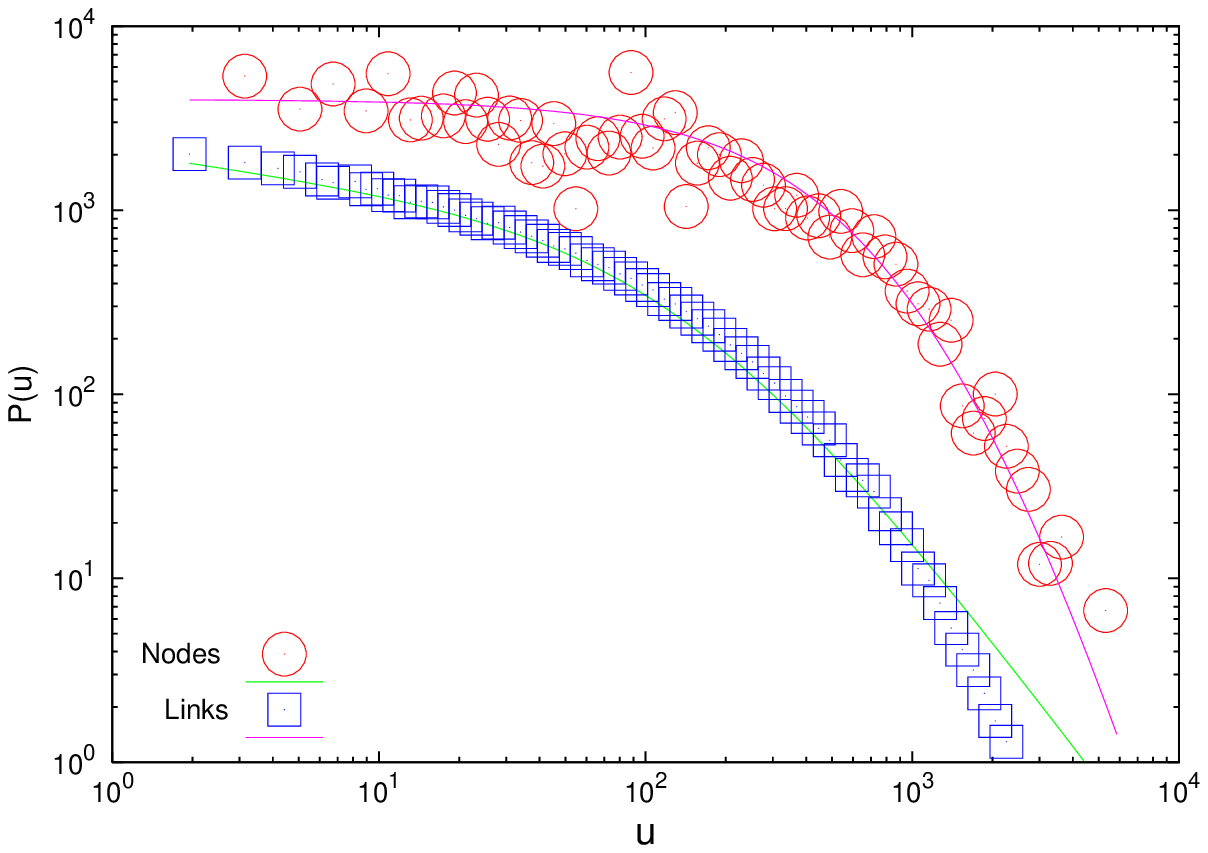}}\\ 
\end{tabular}
\caption{(left) Cumulative distribution of in-coming and out-going links for grown modular network. (right) Distribution of betwenness-centrality for nodes and links in the gene-expression network of Yeast. }
\label{fig-pq}
\end{figure}

At each  time step $t$ we add a new node $i$ and $M$ new links. With probability $P_{o}$ a new group is started. The added node is assigned current group index. (First node belong to the first group.)
The group index plays a crucial role in linking the node to the rest of the network. The links are created by attaching the added node inside the group, with probability ${\tilde{\alpha}}$, or else rewiring within the entire network. The target node \textit{k} is selected preferentially with respect to the current situation in the group, which determines the linking probability $p_{in}(k,t)$. Similarly, the node which rewires the link \textit{n} is selected according to its current number of outgoing links, which determines the probability $p_{out}(n,t)$:
 \begin{equation}p_{in}(k,t)=\frac{M\alpha+q_{in}(k,t)}{tM\alpha+L_{g_{k}}(t)} \;  ,\quad p_{out}(n,t)=\frac{M\alpha+q_{out}(n,t)}{tM(\alpha+1)} \;  ,
\label{pin-pout} 
\end{equation} 
where $q_{in}(k,t)$ and $q_{out}(n,t)$ are in- or out-degrees of respective nodes at time step $t$, $tM(\alpha +1)$ is number of links in whole network, while $L_{g_{k}}(t)$ is number of link between nodes in a group of node $k$. It is assumed that $q_{in}(i,i)$=$q_{out}(i,i)=0$.
Suggested rules of linking insure existence of modules in the network. Each group has  a central node, hub, in terms of in-degree connectivity, and a set of nodes along which it is connected with the other groups.  The  number of groups $G$ in the network depends on number of nodes, \textit{N}, and the parameter \textit{$P_{o}$} as  $G \sim NP_{o}$. Some emergent modular structures are shown in Fig.\ \ref{fig-graphs-clusters}. For the purpose of this work we only mention that the networks grown using the above rules are scale-free with both in-coming and out-going links distributed as 
($\kappa$="in" or "out"):
 \begin{equation}P(q_{\kappa})\sim q_{\kappa}^{-\tau_{\kappa}} \;  .  
\end{equation}
  The scaling exponents $\tau_{in}$ and $\tau_{out}$ vary with the parameters $\alpha$ and \textit{$P_{o}$}. In Fig.\ \ref{fig-pq} we show cumulative distribution of in- and out-links in the case N=25000 nodes, M=5, ${\tilde{\alpha}}=\alpha=0.9$ and number of groups $G=6$. The slopes are $\tau_{in}-1=1.616\pm0.006$ and $\tau_{out}-1=7.6\pm0.3$.

These networks with controlled modularity will be considered in the next Section to test the maximum-likelihood algorithm  for finding subgraphs. In addition, we will apply the method on a gene network, in which the modular structure is not known. 
The  network is based on the empirical data of gene expressions  for a set of 1216 cell-cycle genes of  yeast measured at several points along the cell-cycle, selected from \cite{Y_data}. The pairs of genes are connected with weighted links according to their expression correlation coefficient. In such network, for the correlations exceeding a critical value $W_{0c}$ a percolation-like transition occurs where functionally related clusters of genes join the giant cluster  \cite{Jelena}. However, below that point the network is too dense and separation of the modules becomes difficult. The topological betweenness-centrality measures for both, nodes and links in the gene network shown in Fig.\ \ref{fig-pq} (right), exhibit broad distribution, suggesting a nontrivial topology of the network. 

\section{Maximum Likelihood Methods for Weighted Graphs}
The method is based on a mixture model and numerical technique known as the expectation-maximization algorithm. We first describe the basic idea for unweighted networks, \cite{MLM}, and then generalize the method for weighted graphs.

\subsection{Theoretical Background}
 
A network of $N$ nodes, directed or undirected, is represented mathematically by an adjacency matrix $\mathbf{A}$ with $N \times N$ elements. The elements $A_{ij}$ as (1,0), represent the presence/absence of a link between nodes. 
The idea is to construct a {\it mixture model} network partitioned into $G$ groups, where members of the groups are similar in some sense and the numbers $g_{i}$ denote the group to which vertex $i$ belongs \cite{MLM}. Group memberships are unknown, they are commonly referred as "hidden" data.  The basic idea is to vary parameters of a suitable mixture model to find the best fit to the observed network. 

The model parameters are: $\theta_{ri}$, defined as the probability that a link from some node in group $r$ connects to node $i$, and $\pi_{r}$,  representing the  probability that a randomly chosen vertex falls in group $r$. The normalization conditions 
\begin{equation}
\sum_{r}{\pi_{r}}=1 \;  ,\quad\sum_{i}{\theta_{ri}}=1 \;  ,
\label{norms} 
\end{equation} 
are required. The parameters can be estimated by the maximum likelihood criterion using expectation-maximization algorithm. In the present case the problem reduces to maximization of the likelihood $Pr(A,g \vert \pi,\theta)$ with respect to $\pi$ and $\theta$. Using the factorization rule, we can write $Pr(A,g \vert \pi,\theta)$ in the following form:
\begin{equation}
 Pr(A,g \vert \pi,\theta)=  Pr(A \vert g, \pi,\theta)Pr(g \vert \pi,\theta) \;  ,
\end{equation} where
\begin{equation}
Pr(A \vert g, \pi,\theta)=\prod_{ij}{\theta_{g_{i}j}^{A_{ij}}} \;  ,\quad Pr(g \vert \pi,\theta)=\prod_{i}{\pi_{g_{i}}} \;  .
\end{equation} Combining Eqs. (4) and (5) we obtain
\begin{equation}
 Pr(A,g \vert \pi,\theta)= \prod_{i}{\pi_{g_{i}}\prod_{j}{\theta_{g_{i},j}^{A_{ij}}}} \;  .
\end{equation}

It is common to use the logarithm of likelihood instead of the likelihood itself \cite{MLM}. In addition, averaging of the log-likelihood over distribution of group memberships $g$ is necessary with   
the distribution $Pr(g\vert A,\pi,\theta)$, leading to  
\begin{eqnarray}
\overline{L}&=&\sum^{G}_{g_{1}}\cdots\sum_{g_{n}}^{G}{Pr(g \vert A,\pi,\theta)\sum_{i}{[ln\pi_{g_{i}}+\sum_{j}{A_{ij}ln\theta_{g_{i},j}}]}}\\\nonumber&=&\sum_{ir}{q_{ir}[ln\pi_{r}+\sum_{j}A_{ij}ln\theta_{g_{i},j}]} \;  ,
\end{eqnarray}
where $q_{ir}=Pr(g_{i}=r\vert A,\pi,\theta)$ represents the probability that node $i$ belongs to group $r$. Using again the factorization rule for $Pr(A,g_{i}=r\vert \pi,\theta)$ in the case when $A$ represents the missing data we find the expression for $q_{ir}$

\begin{equation}
q_{ir}=Pr(g_{i}=r\vert A,\pi,\theta)=\dfrac{Pr(A,g_{i}=r\vert \pi,\theta)}{Pr(A\vert \pi,\theta)}=\dfrac{\pi_{r}\prod_{j}{\theta_{rj}^{A_{ij}}}}{\sum_{s}{\pi_{s}\prod_{j}{\theta_{sj}^{A_{ij}}}}} \;  . \label{qir}
\end{equation}

Now we can use  $q_{ir}$ given by the Eq. (\ref{qir}) to evaluate the expected value of log-likelihood and to find  $\pi_{r}$ and $\theta_{ri}$ which maximize it. The maximization can be carried out analytically. Using the method of Lagrange multipliers to enforce the normalization conditions in (\ref{norms}), we find relations for the parameters $\pi_{r}$ and $\theta_{ri}$:
\begin{equation}
\pi_{r}=\dfrac{\sum_{i}{q_{ir}}}{n} \;  , \hspace*{0.5cm}\theta_{ri}=\dfrac{\sum_{j}{A_{ji}q_{jr}}}{\sum_{j}{q_{out}(j)q_{jr}}} \;  .
\label{pi-theta}
\end{equation}
In the numerical implementation, starting from an initial partitioning and iterating the Eqs.\ (\ref{qir}) and (\ref{pi-theta}) towards convergence, we determine the modular structure, which is  defined by quantities $q_{ir}$. In practice, the runtime to the convergence depends on the number of nodes and number of groups, $G$.

In practical calculations,  the  clustering algorithm converges to one out of many local maxima,  which is sensitive to the initial conditions. Hence the choice of initial values of the model parameters  is a not trivial step. The obvious unbiased choice of the initial  values is the symmetric point  with $\pi_{r}^{0}=\dfrac{1}{c}$ and $\theta_{ri}^{0}=\dfrac{1}{N}$, which is consistent with normalization conditions for numbers $\pi^{0}$ and $\theta^{0}$. Unfortunately, this represents a trivial fixed point of the iterations. Instead, we find that the starting values that are perturbed randomly at small distance from the fixed point are better in terms of convergence of algorithm to right maxima of expected value of log-likelihood. After initialization of the parameters, we compute $q_{ir}^{0}$ using Eq. (\ref{qir}) and then that $\pi_{r}^{1}$ and $\theta_{ri}^{1}$ according to Eq.\ (\ref{pi-theta}), etc. After some number of iterative steps, the algorithm will converge to a local maximum of the likelihood. In order to find the global maximum, it is recommendable to perform several runs with different initial conditions.

\hspace*{0.3cm}We test the the MLM algorithm on the networks grown by the model represented in Section 2,  for a wide range values of parameters $M$, $P_{0}$, $\alpha$ and ${\tilde{\alpha}}$. As we expected, the  algorithm works well on networks in which the modules with respect to connectivity---most of the links are between vertices inside the group---are well defined, which is the case for large parameter ${\tilde{\alpha}}$. The partitions are shown in Fig.\ \ref{fig-graphs-clusters}: Clusters of the network found by the algorithm corresponds perfectly to the division derived from the growth.  Group membership suggested by the algorithm and the original one are the same for the 98\% of the nodes in the network. The algorithm even finds the ``connector'' nodes, that play a special role in each group. 
However, it is less efficient for the networks with large density of links between groups, as for instance for  networks with ${\tilde{\alpha}} \approx 0.6$ or lower. Some other observations: Number and size of different groups do not affect the efficiency of the algorithm. Size of the network and sparseness of the network affect the convergence time. As a weak point, the number of groups $G$ is an input parameter.

\subsection{Generalization of the Algorithm for Weighted Networks}

The topological structure of networks, usually expressed  by the presence or absence of links, can be considerably altered when links or nodes aquire different weights.  
Here we modify the algorithm  presented in Section 3.1 in order to take into account {\it weighted networks with modular structure}. 
The main idea is based on the fact that a weighed link between a pair of nodes on the  network can be considered as {\it multiple links} between that pair of nodes. 
Then  a straightforward generalization of the MLM  is to apply the mixture model and expectation-maximization algorithm described above to 
the  {\it multigraph} constructed with the appropriate number of links between pairs of nodes. The quantities to be considered are:
  $W_{ij}$ measured matrix of weights, $g_{i}$ missing data, and model parameters \{$\pi_{r}$,$\theta_{ri}$\}. Following the same steps as above leads to the following expressions which are relevant for the algorithm:
\begin{equation}
q_{ir}=\dfrac{\pi_{r}\prod_{j}{\theta_{rj}^{W_{ij}}}}{\sum_{s}{\pi_{s}\prod_{j}{\theta_{sj}^{W_{ij}}}}} \;  ,
\label{qir_w}
\end{equation}

\begin{equation}
\pi_{r}=\dfrac{\sum_{i}{q_{ir}}}{n}\;  , \hspace*{0.5cm}\theta_{ri}=\dfrac{\sum_{j}{W_{ji}q_{jr}}}{\sum_{j}{l_{j}q_{jr}}} \;  ,
\label{pi-theta_w}
\end{equation} 
where $l_{j}=\sum_{i}{W_{ji}}$ is the summation over all weights of links emanating from the node \textit{j}. The implementation of the algorithm and the choice of initial values of parameters are illustrated in previous chapter. Although the formal analogy between the expressions in Eqs.\ (\ref{pi-theta}) and (\ref{pi-theta_w}) and also between Eqs.\ (\ref{qir}) and (\ref{qir_w}) occurs with $W_{ij}\to A_{ij}$, the important difference occurs in the quantity $l_j$ in Eq.\ (\ref{pi-theta_w}), which is a measure of {\it strength} of node rather than its connectivity. Therefore, within the weighted algorithm, nodes of same strength appear to belong to the same community. The weighted communities may have important effects in the dynamics, as for instance, the clusters of nodes with same strength tend to synchronize at the same time scale.
Such properties of the weighted networks remain elusive for the classical community structure analysis based on max-min theorem, mentioned in the introduction.   

\begin{figure}
\centering
\begin{tabular}{cc} 
\includegraphics[height=0.3\textheight,width=0.3\textheight]{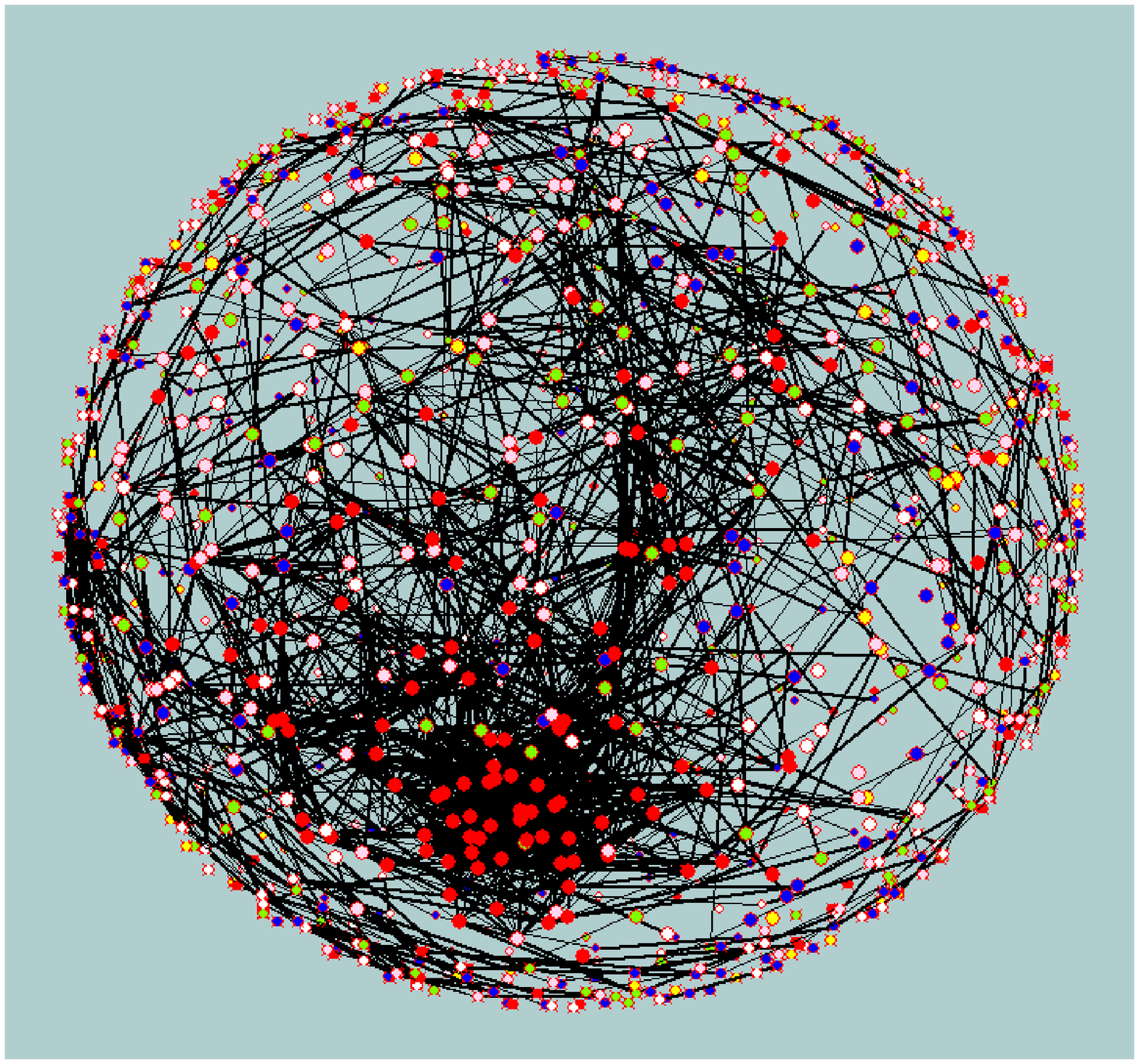}&
\includegraphics[height=0.3\textheight,width=0.3\textheight]{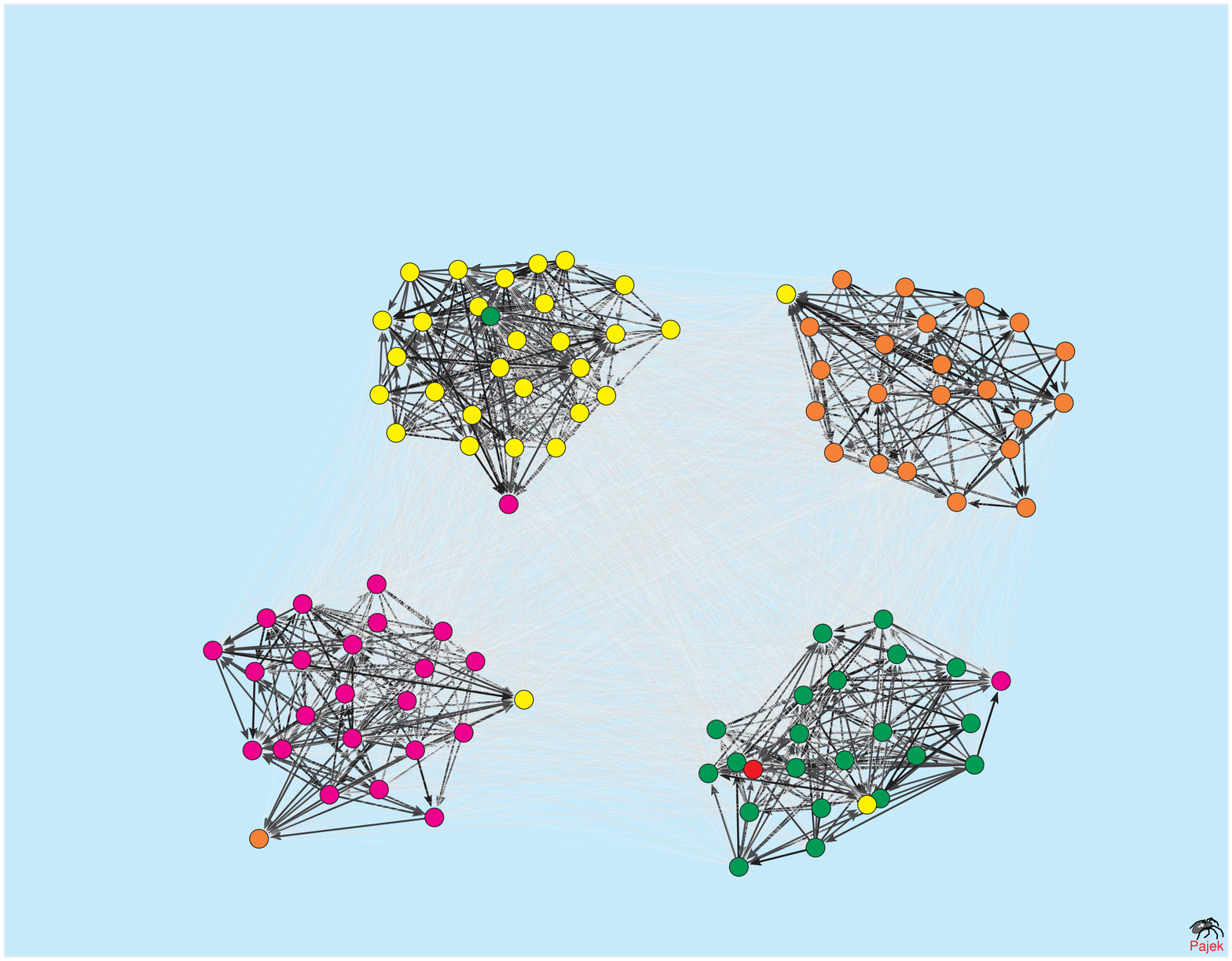}\\ 
\end{tabular}
\caption{Weighted clusters found by the extended ML algorithm in the correlation network of gene expressions of yeast (left),  and in the weighted random graph (right). }
\label{fig-weightednets}
\end{figure}

\hspace*{0.3cm}In the remaining part of the Section we apply the algorithm to two weighted networks: first we demonstrate how it finds weighted subgraphs on a computer generated random graph with large density of links, and then on a gene expression network of yeast generated from the empirical expression data. The results are given in Fig.\ \ref{fig-weightednets}. 
A random graph of $N=100$ nodes with a link between each pair of nodes occurring with probability $p=0.5$ is generated. In this homogeneously connected graph we create $G=4$ groups and assign different weights to links in each group of nodes. In the network of yeast gene expressions, as described in Section 2, the weights of links appear through the correlation coefficient of the gene expressions.  As the Fig.\ \ref{fig-weightednets} shows, the algorithm retrieves accurately four {\it a-priori} known weighted groups of nodes on the random graph. Similarly, in the case of gene networks, we tried several partitiones with the potentially different number of groups. One such partition with $G=5$ groups is shown in Fig.\ \ref{fig-weightednets}. 
Each weighted group  on the gene network actually represents a set of genes which are closely co-expressed during the entire cell cycle.   

\section{Conclusions}
We have extended the maximum-likelihood-method of community analysis to incorporate multigraphs (wMLM) and  analysed several types of  networks with mesoscopic inhomogeneity. Our results show that the extended wMLM can be efficiently applied to search for variety of subgraphs from a clear topological inhomogeneity with network-of-networks structure, on one end, to hidden subgraphs of node with the same strength on the other. \\
\\
{\bf Acknowledgments:} Research supported in part by national projects P1-0044 (Slovenia) and OI141035 (Serbia), bilateral project BI-RS/08-09-047 and COST-STSM-P10-02987 mission. The numerical results were obtained on the AEGIS e-Infrastructure, supported in part by EU FP6 projects EGEE-II, SEE-GRID-2, and CX-CMCS.

\end{document}